\documentclass[conference]{IEEEtran}
\IEEEoverridecommandlockouts
\usepackage{cite}
\usepackage{amsmath,amssymb,amsfonts}
\usepackage{algorithmic}
\usepackage{graphicx}
\usepackage{textcomp}
\usepackage{xcolor}
\usepackage{booktabs}

\usepackage{caption}
\usepackage{subcaption}
\usepackage{enumitem}

\def\BibTeX{{\rm B\kern-.05em{\sc i\kern-.025em b}\kern-.08em
    T\kern-.1667em\lower.7ex\hbox{E}\kern-.125emX}}
\begin{document}

\title{Universal Auto-encoder Framework for MIMO CSI Feedback
}

\author{\IEEEauthorblockN{Jinhyun So}
\IEEEauthorblockA{
\textit{Samsung Semiconductor Inc.}\\
San Diego, USA \\
jinhyun.so@samsung.com}
\and
\IEEEauthorblockN{Hyukjoon Kwon}
\IEEEauthorblockA{
\textit{Samsung Semiconductor Inc.}\\
San Diego, USA \\
hyukjoon.k@samsung.com}
\vspace{-1.0cm}
}
% \and
% \and
% \and
% \and

\maketitle
\vspace{-1.0cm}
\begin{abstract}
Existing auto-encoder (AE)-based channel state information (CSI) frameworks have focused on a specific configuration of user equipment (UE) and base station (BS), and thus the input and output sizes of the AE are fixed.
% % 
However, in the real-world scenario, the input and output sizes may vary depending on the number of antennas of the BS and UE and the allocated resource block in the frequency dimension.
A naive approach to support the different input and output sizes is to use multiple AE models, which is impractical for the UE due to the limited HW resources.
In this paper, we propose a universal AE framework that can support different input sizes and multiple compression ratios.
The proposed AE framework significantly reduces the HW complexity while providing comparable performance in terms of compression ratio-distortion trade-off compared to the naive and state-of-the-art approaches.

\end{abstract}


\section{Introduction}
In a MIMO system, real-time channel state information (CSI) at the base station (BS) plays an important role in taking advantage of advanced multi-input multi-output (MIMO) techniques. However, in frequency division duplex (FDD) systems, the user equipment (UE) needs to estimate the downlink CSI based on reference signals and send it back to the BS. The communication overhead for CSI feedback becomes one of the major challenges for Massive MIMO in FDD systems, which presents a non-trivial trade-off between CSI distortion and feedback rate. To address this challenge, several methods using compressed sensing (CS) and codebooks have been introduced and applied to LTE and New Radio (5G). The complexity of codebook design and exploiting the sparsity of CSI increases exponentially with the number of transmit and receive antennas, making it impractical in MIMO systems.

\begin{figure}[!t]
\centering
\includegraphics[width=0.95\linewidth]{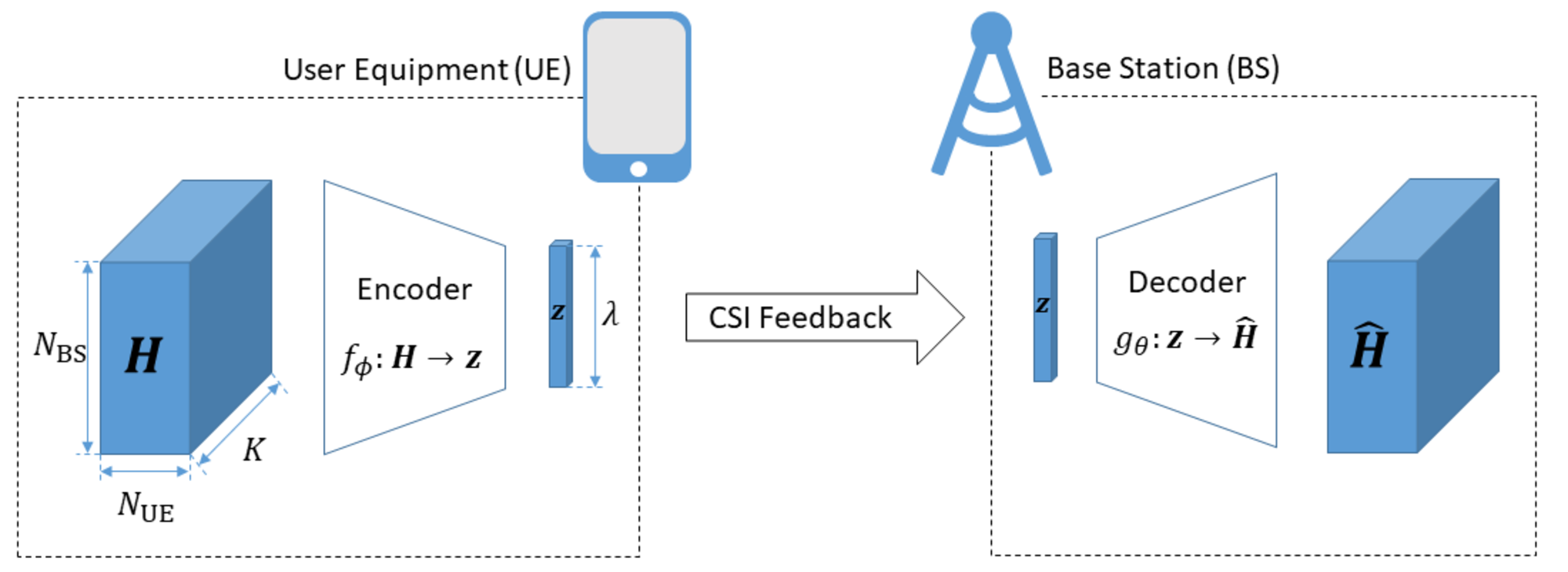}
\vspace{-0.1cm}
\caption{Auto-encoder (AE) based framework for channel state information (CSI) feedback system.}
\label{fig:feedback_system}
\vspace{-0.2cm}
\end{figure}

Recently, machine learning (ML)-based methods for CSI compression have been studied~\cite{guo2022overview, song2021saldr, xu2021transformer, kim2022channel}. Most of the works in this line consider an auto-encoder (AE) architecture consisting of the pair of encoder and decoder, where the encoder takes CSI as input and outputs a latent vector whose size is much smaller than the size of the input. The decoder in AE aims to reconstruct the original CSI based on the latent vector. The goal of AE is to learn a nonlinear manifold of the CSI with a few dimensions, which is well suited to the CSI feedback problem, as shown in Fig.~\ref{fig:feedback_system}. It has been shown that the AE-based methods can provide a better trade-off between distortion and feedback rate than the conventional CS and codebook-based schemes~\cite{guo2022overview, song2021saldr, xu2021transformer, kim2022channel}. 
% As such, 3gpp is considering AI-based CSI feedback enhancement as a study item for Release 18 [5]. 

\begin{figure}[!t]
\centering
\includegraphics[width=0.92\linewidth]{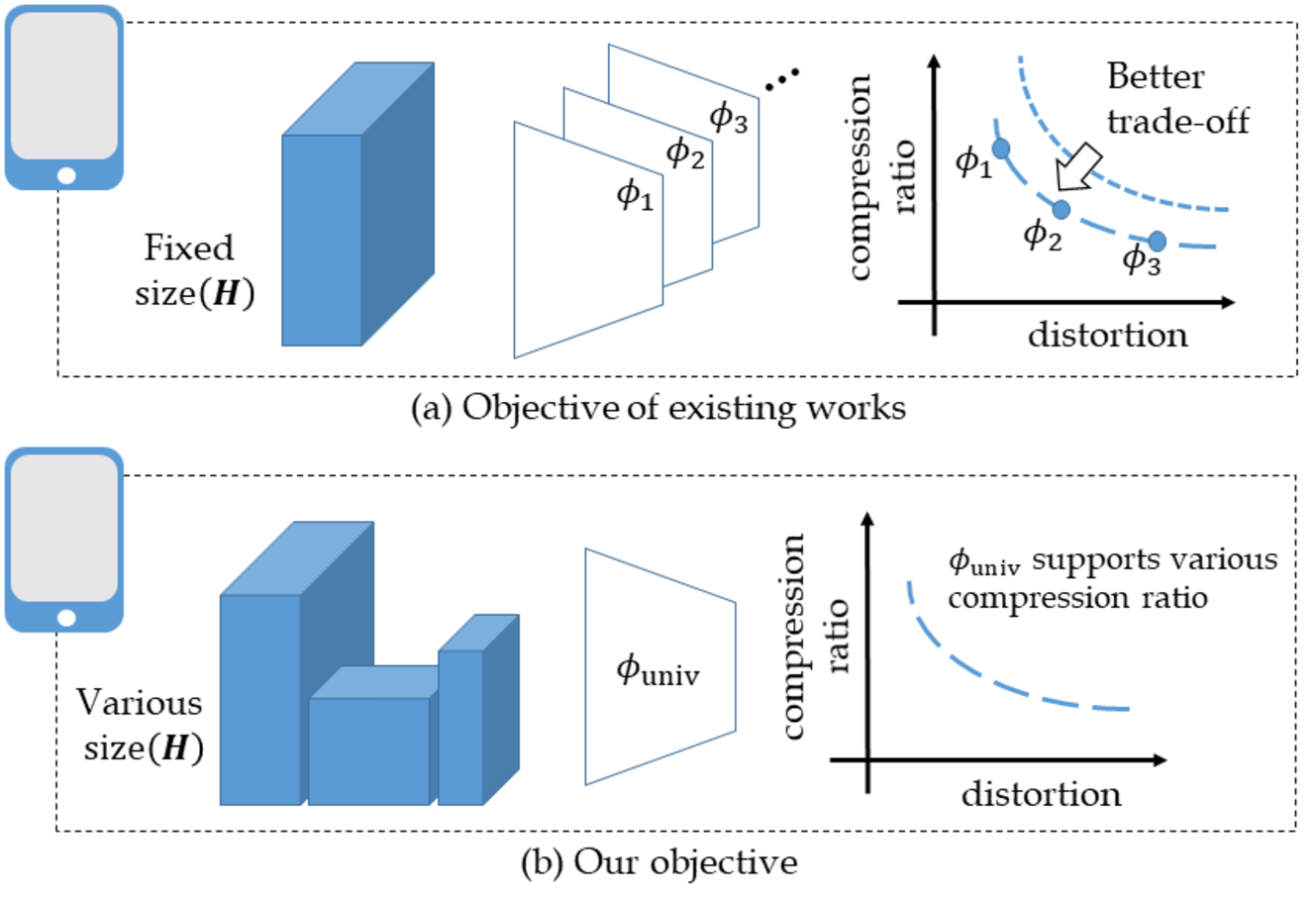}
\vspace{-0.1cm}
\caption{Comparison of objectives of existing works and ours. (a) Existing works aim to design multiple encoders ($\phi_1, \phi_2, \ldots$) providing a better trade-off between compensation ratio and distortion while the input size is fixed. (b) Our work aims to design a universal encoder ($\phi_{\mathtt{univ}}$) which can support various input and latent size to reduce the hardware (HW) complexity of the UE while it shows comparable performance in terms of compression-distortion trade-off. 
}\label{fig:objective}
\vspace{-0.3cm}
\end{figure}

Most of the existing works have focused on designing an efficient AE framework given the specific configuration, where the number of antennas and the resource allocation to the UE in the frequency domain are fixed. However, in the real-world scenario, the number of transmit and receive antennas can vary among BSs and UEs, and the BS dynamically allocates the resource block in the frequency domain to the UE according to the channel quality. Therefore, the input dimension of the encoder may change, and thus the UE may need to use multiple encoders to support the different input sizes. In addition, the UE may also need to support different latent sizes and compression ratios (CRs), as the recent work has shown that further communication overhead reduction in CSI feedback can be achieved by adjusting the latent size according to the channel delay profile~\cite{kim2022learning}. The straightforward way to support the different input and latent sizes is to design multiple AE models, each dedicated to the input-latent-size pair. However, due to the limited hardware (HW) resources in mobile devices, implementing multiple encoders for the UE may be impractical.  

To address this challenge, our objective is to design a universal AE-based CSI feedback framework that can support different input and latent sizes. Fig.~\ref{fig:objective} shows the comparison of the objectives of the existing works and ours. Our proposed framework has two outstanding features:
\begin{itemize}[leftmargin=*, itemsep=1pt]
    \item The encoder framework in the UE is agnostic to the BS configurations such as the number of BS antennas and allocated resource block from the BS.

    \item A single set of ML parameters of the encoder can support multiple compression ratios (CRs). The encoder contains a universal block without additional layers for each CR, and we propose a novel ML training scheme using a masking layer, which makes the output of the universal block contain more important information in the earlier position. 
\end{itemize}

% \textcolor{red}{[ADD CONTRIBUTIONS HERE.]}

\section{System Model}
In this section, we introduce the limited CSI feedback systems for Massive MIMO and provide a standard AE architecture.
The AE captures the salient features of high-dimensional MIMO channels, which significantly reduces the feedback overhead. 
We consider a massive MIMO orthogonal frequency division multiplexing (OFDM) system where a single user (UE) and a single BS have $N_{UE}$ and $N_{BS}$, respectively. The BS transmits OFDM transmission with $N_s$ data streams over $K$ subcarriers. The received signal on the $k$th subcarrier can be expressed as
\begin{equation}
    \mathbf{y}_k=\mathbf{H}_k^H \mathbf{V}_k \mathbf{x}_k + \mathbf{n}_k,
\end{equation}
where $\mathbf{H}_k\in \mathbb{C}^{N_{BS}\times N_{UE}}$, $\mathbf{V}_k\in \mathbb{C}^{N_{BS} \times N_s}$, $\mathbf{x}_k\in \mathbb{C}^{N_s}$, and $\mathbf{n}_k\in\mathbb{C}^{N_{UE}}$ denotes the channel matrix in frequency domain, precoding matrix at the BS, downlink transmitted data symbol, and additive white Gaussian noise on the $k$-th subcarrier, respectively. 
Let $\mathbf{H}=\left\{\textnormal{real}(\{\mathbf{H}_1,\mathbf{H}_2,…,\mathbf{H}_K \}),\textnormal{imag}(\{\mathbf{H}_1,\mathbf{H}_2,…,\mathbf{H}_K \})\right\}\in \mathbb{R}^{2\times K \times N_{BS} \times N_{UE}}$ be a tensor representing the entire CSI stacked by the channel matrices on all subcarriers 
while $\textnormal{real}(\cdot)$ and $\textnormal{imag}(\cdot)$ denote the real and imaginary part of the input, respectively. 
We note that resolution in the frequency dimension can vary according to the granularity of AI-based CSI feedback which can be defined by the network, and in this paper we use a resource block (RB).

Based on the CSI information $\mathbf{H}$, the BS designs the precoding matrixes to improve the spectral efficiency with various MIMO techniques including eliminating the inter-user interference or beamforming. For doing so, the UE estimates and sends $\mathbf{H}$ to the BS, but the total number of feedback parameters in FDD massive MIMO system is linearly increasing with respect to $N_{UE}$,$N_{BS}$, and $K$, which is too large in feedback links. 
To reduce the feedback overhead, the UE can extract the most salient features of the CSI information H by utilizing the AE which employs the pair of encoder and decoder to compress and reconstruct the CSI, respectively. The encoder carries out the compression as
\begin{equation}
    \mathbf{z} = f_{\phi}(\mathbf{H}),
\end{equation}
where $\mathbf{z}\in\mathbb{R}^{\lambda}$ is a latent vector of size $\lambda$ and $f_\phi(\cdot)$ denotes the compression function with parameters $\phi$. Then the compression ratio is defined as the ratio between the input and output dimension of the encoder function $f_{\phi}$, which can be expressed by
\begin{equation}
    CR_{f_\phi} \triangleq \frac{\mathtt{size}(\text{output of }f_\phi)}{\mathtt{size}(\text{input of }f_\phi)} = \frac{\lambda}{2 K N_{UE} N_{BS}},
\end{equation}
% test
where $\mathtt{size}$$( \cdot )$ denotes the number of elements.
Note that in this paper, we assume the unlimited representations of continuous features over the latent space. 
We remain joint optimization of the encoder and quantization to enable discrete representation of the latent space as one of future directions. 
The UE sends the compressed version of CSI, the latent vector $\mathbf{z}$, to the BS, which significantly reduces the feedback overhead when $\lambda \ll 2K N_{UE} N_{BS}$. 
Upon receiving $\mathbf{z}$, the BS reconstruct the CSI information by carrying out the decoder, which can be expressed by
\begin{equation}
    \hat{\mathbf{H}} = g_{\theta}(\mathbf{z}), 
\end{equation}
where $g_{\theta} (\cdot)$ denotes the reconstruction function with a set of parameters $\theta$, and $\hat{\mathbf{H}}$ is the reconstructed CSI tensor which have the same dimensionality as $\mathbf{H}$. 

As described in the introduction, CSI tensor $\mathbf{H}$ can have various dimensionality according to the UE and BS antenna configuration, and resource allocation over the frequency dimension. 
Various feedback overhead (and hence the latent size $\lambda$) can be configured to optimize the trade-off between distortion and communication overhead \cite{kim2022learning}. Let $\Lambda=\{\lambda_1,\lambda_2,…,\lambda_{max} \}$ be a set of latent sizes that we need to support and $\lambda_{max}$ is the maximum value of the latent vector size. Let $\mathcal{D}=\{D_1,D_2,…\}$ be a set of distributions of CSI tensor, and each distribution may have different size. Straightforward approach to support all pairs in $\Lambda\times\mathcal{D}$ is that for each pair of $(\lambda_i,D_j )\in\Lambda\times\mathcal{D}$, the dedicated pair of encoder and decoder is trained to minimize the reconstruction loss, which can be expressed by
\begin{equation}\label{eq:conv_obj}
    \phi_{\lambda_i, D_j}^{opt}, \theta_{\lambda_i, D_j}^{opt} 
    = \underset{\phi, \theta}{\arg\max} \mathbb{E}_{\mathbf{H}\sim D_j}\left[ L\left( \mathbf{H}, g_{\theta_{\lambda_i}} \left( f_{\phi_{\lambda_i}}(\mathbf{H})\right) \right)  \right],
\end{equation}
where $L(\cdot)$ denotes the loss function. In the worst case, the number of set of the AE pairs becomes $|D|\times |\lambda|$, which can be impractical to be implemented in the UE.
To address this challenge, our objective is to design a universal encoder in the UE which minimizes the total loss function, which can be expressed by
\begin{align}
    &\phi_{\texttt{univ}}^{opt}, \{\theta_{\lambda_i}^{opt}\}_{\lambda\in\Lambda} 
      \label{eq:our_obj} \\
    &=\underset{\phi, \{\theta_\lambda\}_{\lambda\in\Lambda}}{\arg\max} \sum_{D \in \mathcal{D}} \sum_{\lambda\in\Lambda} \mathbb{E}_{\mathbf{H}\sim D}\left[ L\left( p_D(\mathbf{H}), g_{\theta_{\lambda}}\left( f_{\phi}(p_D(\mathbf{H}))\right) \right)  \right] \notag,
\end{align}
where $p_D (\cdot)$ is a pre-processing function to make the output belong to the same space $\mathcal{H}$ while the input size of function $p_D$ depends on $D$. The role of $p_D$ is to make the input of $\phi_\mathtt{univ}$ have the same size.  
Again, our goal is to find the optimal universal encoder $f_\phi:\mathcal{H}\rightarrow \mathbb{R}^{\lambda_\mathtt{max}}$ from \eqref{eq:our_obj} and $p_D$ such that the AE framework can support arbitrary pair of the input and latent size $(\mathtt{size}(D),\lambda)$ for all $D\in\mathcal{D},\lambda\in\Lambda$. The BS may have the different decoder model $g_{\theta_{\lambda}}$ for each $\lambda\in\Lambda$.

\section{Universal Auto-encoder Framework}
In this section, we investigate how to design AE-based CSI feedback framework which can support various configurations (i.e., various size of the CSI tensor) and multiple compression ratios (i.e., various size of the latent vector) with the limited HW resource of the UE. 

\subsection{Input Space Generalization}\label{subsec:input_generalization}
At a high-level, input space generalization is composed of two components, 1) partition of input CSI tensor over antenna domains and 2) zero-padding in frequency domain. Partitioned and zero-padded inputs are compressed by a single encoder. The advantage of this partition-based structure is two-folds:
\begin{itemize}[leftmargin=*, itemsep=1pt]
    \item (Supporting arbitrary input size) As the single (and same) encoder is applied to each antenna element, it can support arbitrary number of UE and BS antennas,
    \item (Reducing HW complexity) As the input size of the encoder is reduced from the partition (i.e., $\mathbf{H}\in \mathbb{R}^{2\times K\times N_{BS}\times N_{UE}}\rightarrow \mathbf{h}^{\text{part}}\in \mathbb{R}^{2\times K\times1\times1}$), the number of learnable parameters in the encoder can be significantly reduced. The details will be explained later, but note that the number of parameters of a simple multiple layer perceptron (MLP) structure is reduced from 24 millions (M) to 37 thousands (K).
\end{itemize}
Then, the natural question is that as the reduced size of encoder is repeatedly applied to different antenna element, what is the performance degradation from the partitioning? Surprisingly, the performance degradation in terms of BLER is negligible while the number of parameters is reduced from 24M to 37K. Now, we will explain the details of the input space generalization. 

Compressing the entire CSI tensor $\mathbf{H}\in\mathbb{R}^{2\times K \times N_{BS} \times N_{UE}}$ at once may be intractable to the UE. Let's consider the case where the values of $(N_{BS}$, $N_{UE}, K)$ is up to $(32, 4, 273)$. The problem is that with these maximum values of input size of the AE, the number of parameters of a simple MLP encoder with one hidden layer, which is one of the simplest ML models, is more than 24M. Due to the limited HW resource of the UE, implementing these parameters is not practical. Given the fixed compression ratio, the number of parameters of the AE is quadratic to the input size in general. Therefore, in this subsection, we investigate how to reduce the input size while preserving the reconstruction performance as much as possible.

% \begin{figure}[t]
% \centering
% \includegraphics[width=\linewidth]{figures/input_partitioning.eps}
% \caption{Three settings of partitioning $\mathbf{H}$ to reduce the input size of the AE.}
% \label{fig:input_partitioning}
% \vspace{-0.3cm}
% \end{figure}

\begin{figure}
     \centering
     \begin{subfigure}[b]{0.4\textwidth}
         \centering
         \includegraphics[width=\textwidth]{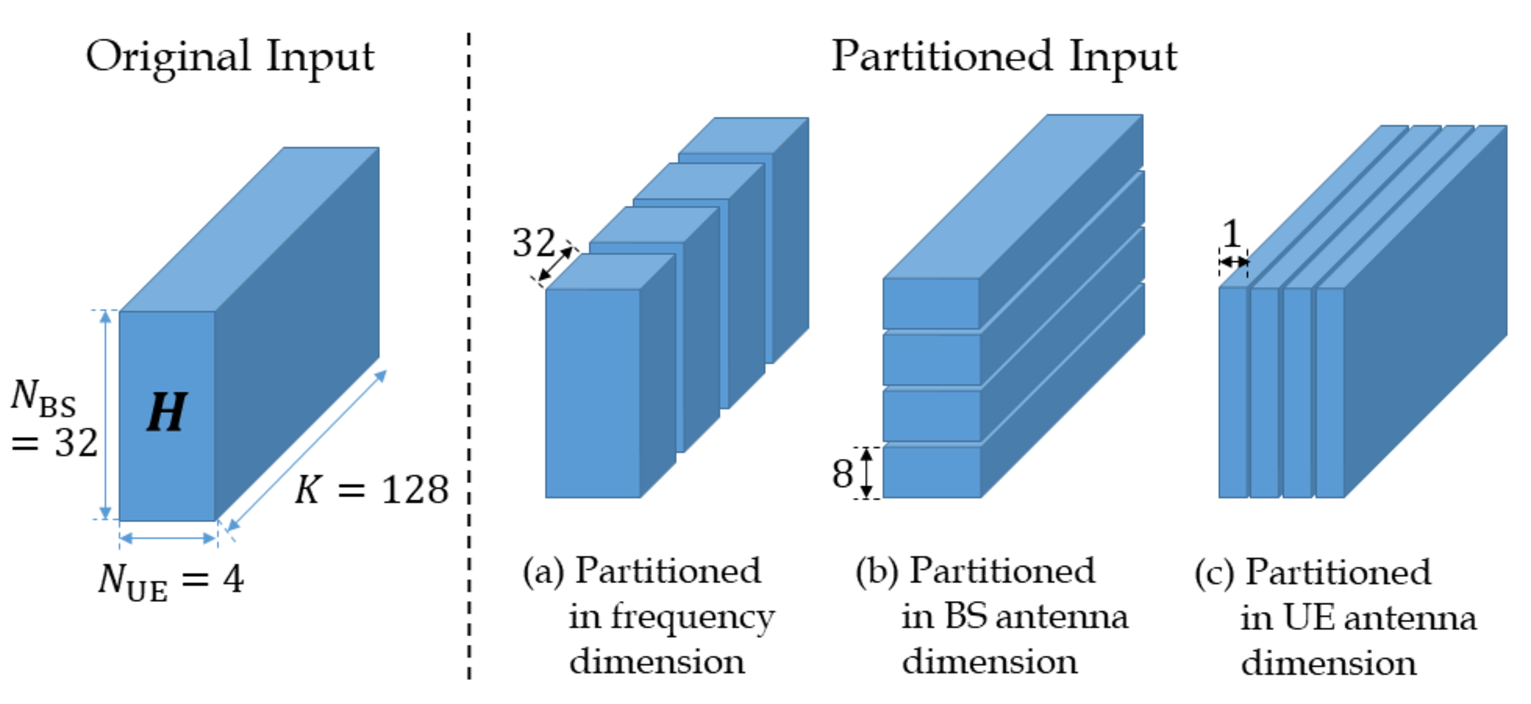}
         \vspace{-0.15cm}
         \caption{Three settings of partitioning $\mathbf{H}$.}
         % \vspace{-0.08cm}
         \label{fig:input_partitioning}
     \end{subfigure}
     \hfill
     \begin{subfigure}[b]{0.32\textwidth}
         \centering
         \includegraphics[width=\textwidth]{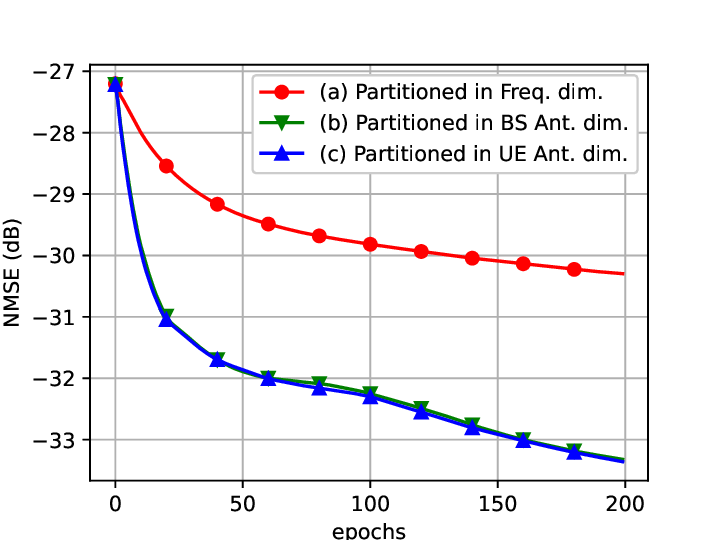}
         \caption{NMSE performance of three settings.}
         \label{fig:NMSE_input_partitioning}
     \end{subfigure}
    \caption{Three settings of partitioning $\mathbf{H}$ to reduce the input size of the AE and their NMSE performance.}
    \label{fig:input_partition_results}
\end{figure}

\begin{figure}[!t]
\vspace{-0.3cm}
\centering
\includegraphics[width=0.55\linewidth]{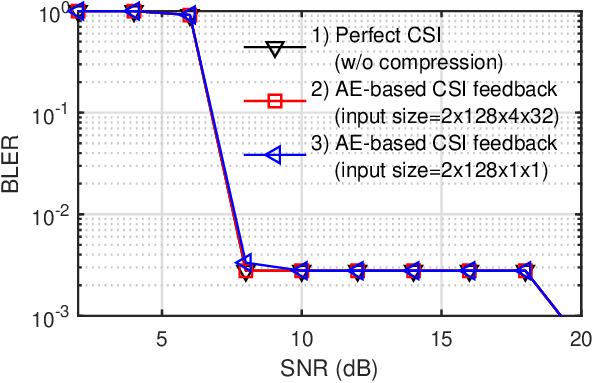}
\vspace{-0.12cm}
\caption{BLER performance of SVD-based beamforming with three types of CSI feedback.}
\label{fig:BLER_input_gen}
\vspace{-0.35cm}
\end{figure}

To reduce the input dimension of the AE, we consider a partition-based approach where the UE partitions $\mathbf{H}$ into multiple parts, each part is compressed by carrying out the encoder, and the UE sends the concatenation of the compressed parts to the BS. Partition can be performed according to the one of dimensions of $\mathbf{H}$, i.e., frequency, BS antenna and UE antenna dimension. To investigate which dimension is \emph{good} to be partitioned, we compare the NMSE performance of three settings that $\mathbf{H}\in\mathbb{R}^{2\times K(=128) \times N_{BS}(=32) \times N_{UE}(=4)}$ is partitioned input $4$ parts with respect to frequency, BS antenna, and UE antenna dimension, respectively. Fig.~\ref{fig:input_partitioning} shows the size of the AE input in three settings. \emph{Good} partition means that the NMSE performance does not degrade even if the same encoder is applied to all parts to save the HW resources.
Fig.~\ref{fig:NMSE_input_partitioning} shows the NMSE performance of the three settings of partitioning. We observe that the UE antenna dimension and the BS antenna dimensions are more robust against performance degradation from partitioning than the frequency dimension. This is because elements in frequency dimension are highly correlated than the other dimensions and the sparsity property over the frequency dimension is good to be compressed. 
% We set low antenna correlation in the simulation for Figure 4, but we believe that even in the high antenna correlation scenario, frequency dimension is better to be compressed, which will be demonstrated later.

To further reduce the input size, we investigate how much the block error rate (BLER) performance is degraded when $\mathbf{H}$ is partitioned such that each part has a single element in the BS and UE antenna dimensions, i.e., $\mathbf{h}^{\mathtt{part}} \in \mathbb{R}^{2×K×1×1}$. 
Fig.~\ref{fig:BLER_input_gen} shows the BLER performance of three cases: the BS performs the singular vector decomposition (SVD)-based beamforming with 1) $\mathbf{H}$ (without compression) and 2) reconstructed CSI matrix  $\hat{\mathbf{H}} = g_\theta (f_\phi (\mathbf{H}))$ where the CSI tensor is not partitioned, and 3) concatenation of reconstructed channel vectors, i.e., $\hat{\mathbf{H}} = \mathtt{concat} \left( \left\{ g_\theta (f_\phi (\mathbf{h}^{\mathtt{part}})) \right\} \right)$. 
Surprisingly, second and third cases have the almost same BLER performance while the number of parameters are significantly reduced by partitioning the input and applying the same AE model to all parts. Note that the number of parameters of the second and third cases are 24M and 37K, respectively.

\begin{figure}[t]
\centering
\vspace{0.1cm}
\includegraphics[width=\linewidth]{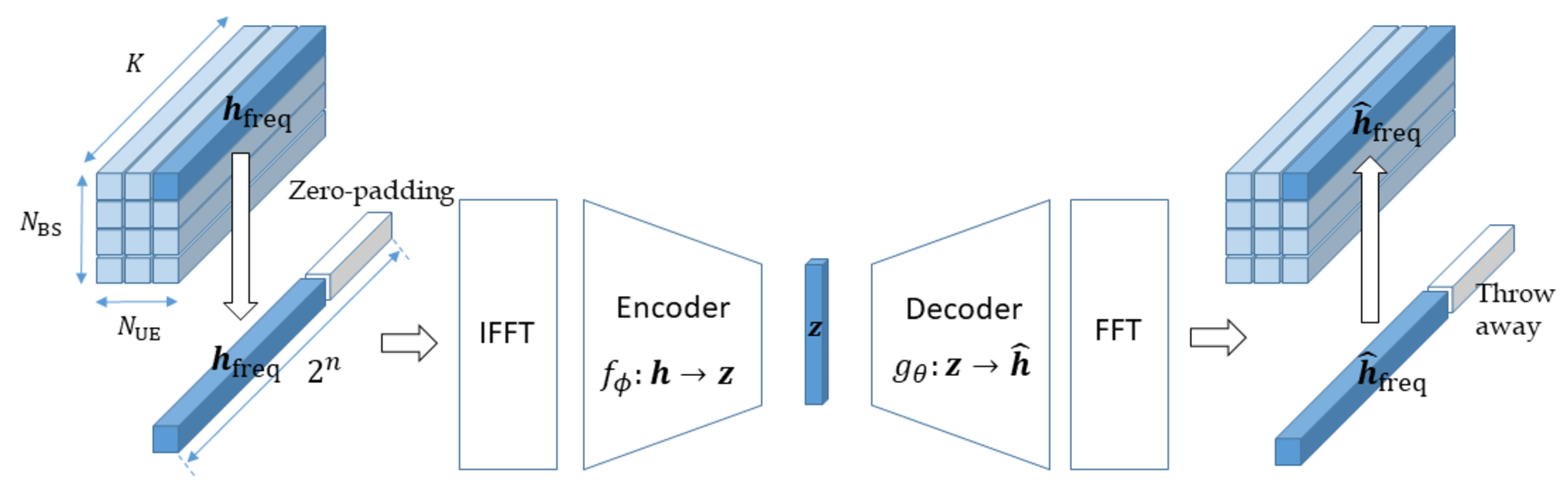}
\caption{AE-based CSI feedback framework with the input space generalization when $K\in(2^{n-1},2^n]$ ($n$ is an integer).}
\label{fig:input_generalization}
\vspace{-0.1cm}
\end{figure}

Now, the AE can be applied to arbitrary number of elements in the UE and BS antenna dimensions as the CSI tensor can be partitioned such that each part has a single element in those dimension and the same AE model can compress all parts without performance degradation. Remaining one is frequency dimension, whose size depends on CSIRS resource allocation and the number of allocated resource block (RB) can be any integer from 1 to 273. 
Zero-padding in frequency domain and applying IFFT may enable a single AE to support entire cases of CSIRS resource allocation. 
In the extreme case, however, to compress a single RB by using the AE whose input size is 273 RB is inefficient in terms of power consumption as 272 elements out of 273 elements in the input of the AE are zero. 
To avoid this inefficiency, we categorize the input size in frequency dimension into 5 cases according to the number of RB(=$K$). 
In each category, $\mathbf{h}^\mathtt{part}$ is zero-padded such that the number of elements in frequency domain has the form of $2^n$ where $n$ is an integer, and then is converted to delay dimension by applying $2^n$-point IFFT. 
To save the power consumption, the UE maintains 5 encoders whose input size in delay dimension is 16, 32, 64, 128, and 256, respectively. 
Table~\ref{tbl:input_category} summarizes the 5 categories with number of RB and IFFT size. Fig.~\ref{fig:input_generalization} shows the overall framework of the CSI compression and reconstruction.

Now, we can view the set $\mathcal{D}$ in \eqref{eq:our_obj} as a set of five distributions $\{D_1, D_2, \ldots,D_5\}$ and $D_i$ is an underlying distribution of the zero-padded and converted (by IFFT) CSI tensor in $i$-th input category. The precoding function $p_{D_i}$ in \eqref{eq:our_obj} is zero-padding to have $2^{i+3}$ elements and applying $2^{i+3}$-point IFFT.

\begin{table}[t!]
\small
\vspace{0.2cm}
\caption{Categories of the input space according to the number of elements in frequency dimension.}
\vspace{-0.2cm}
\label{tbl:input_category}
\begin{center}
% \begin{small}
% \begin{sc}
\begin{tabular}{cccccc}
\toprule
 Category No. & 1 & 2 & 3 & 4 & 5   \\
\midrule
 RB (=$K$)  & $1\sim16$ & $\sim32$ & $\sim64$ & $\sim128$ & $\sim273$   \\
 IFFT size  & $16$      & $32$       & $64$       & $128$       & $256$         \\
\bottomrule
\end{tabular}
% \end{sc}
% \end{small}
\end{center}
\vspace{-0.1cm}
% \vskip -0.1in
\end{table}

\subsection{Latent Space Generalization}\label{subsec:latent_generalization}

In this subsection, we investigate how to design the AE framework which supports multiple compression ratios. Key intuition behind the latent space generalization is that parameters of the encoder are trained such that earlier position in the latent space contains more important information. Then, the encoder can output the elements starting from the earlier position according to the given compression ratio. Before presenting our proposed approach, we introduce two baseline approaches.  

\begin{figure}
     \centering
     \begin{subfigure}[b]{0.4\textwidth}
         \centering
         \includegraphics[width=\textwidth]{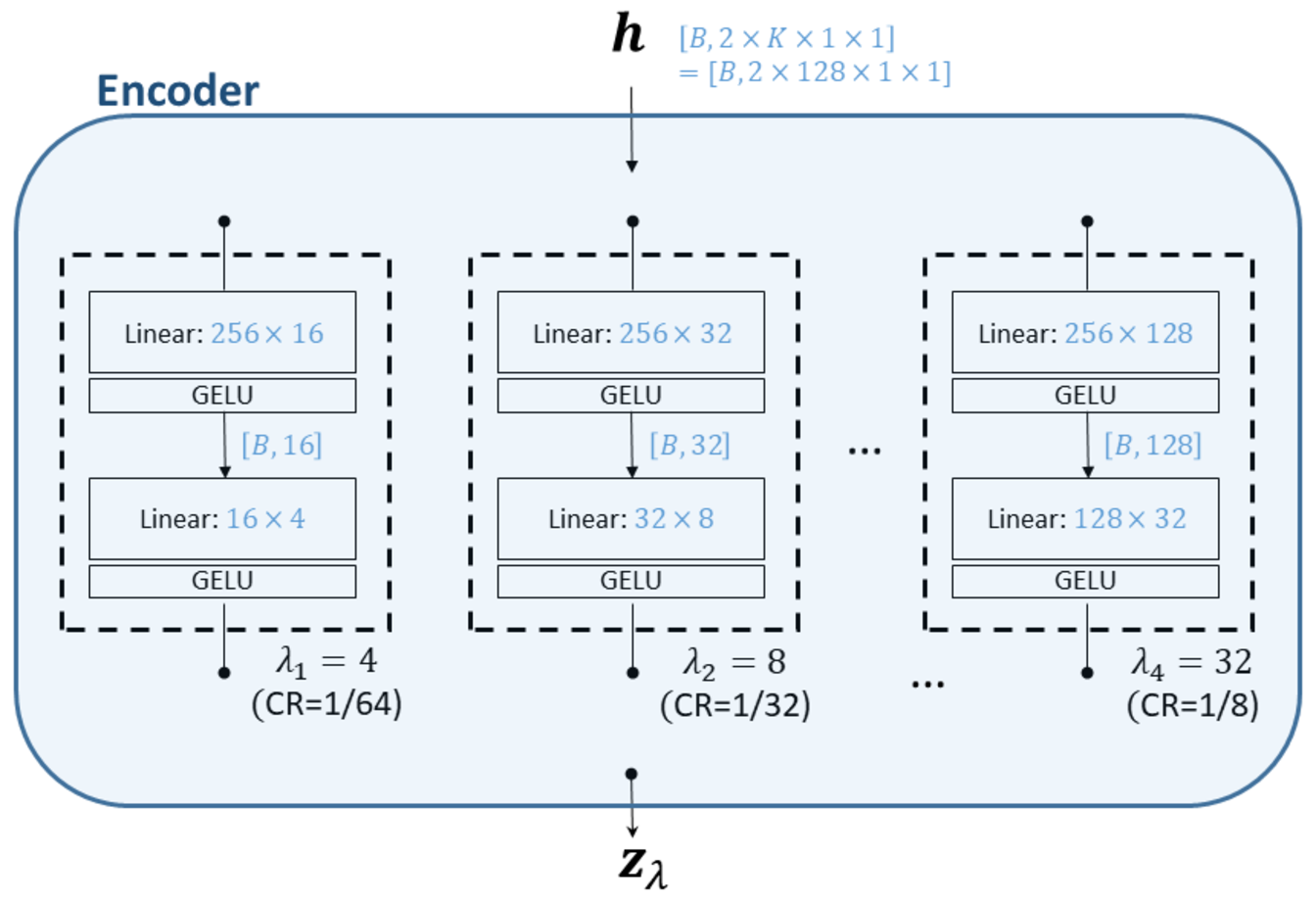}
         \caption{Naive Approach}
         \label{fig:block_approach1}
     \end{subfigure}
     \hfill
     \begin{subfigure}[b]{0.35\textwidth}
         \centering
         \includegraphics[width=\textwidth]{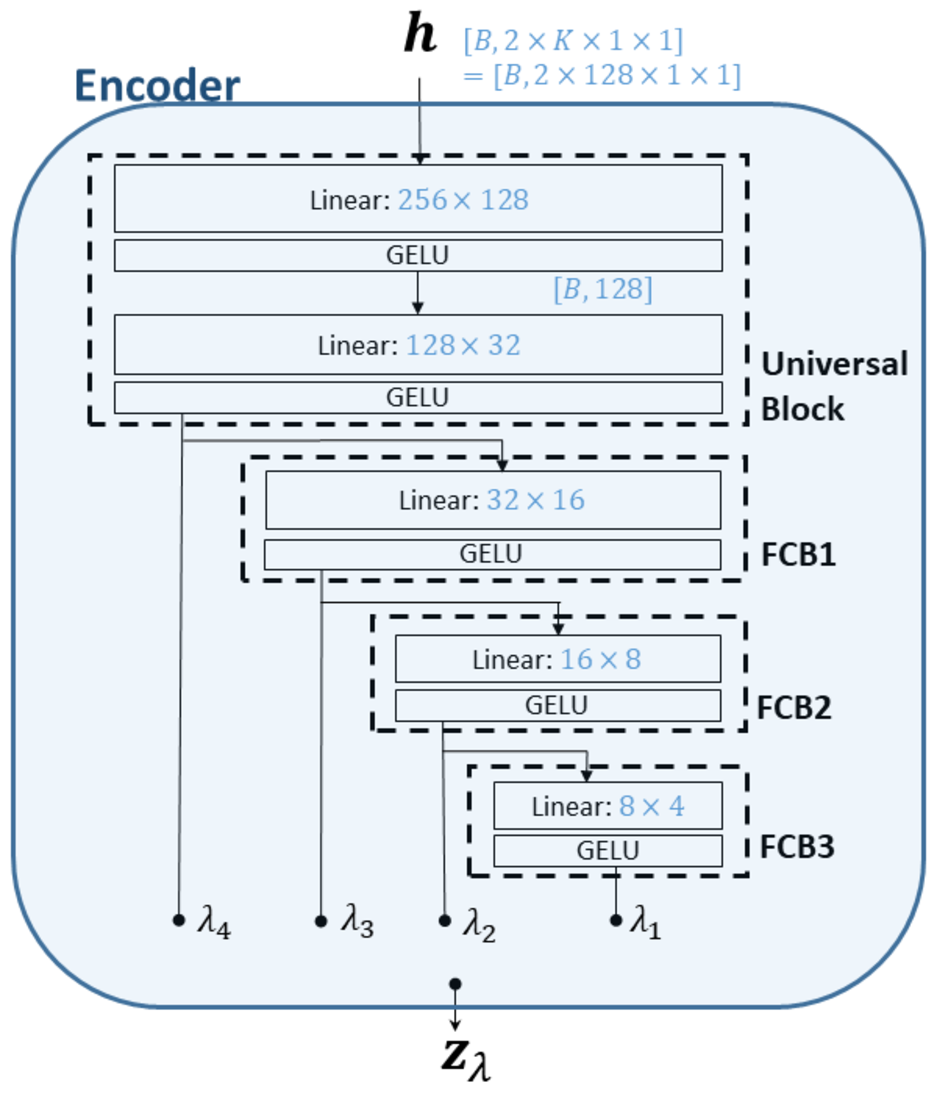}
         \caption{SALDR~\cite{song2021saldr}}
         \label{fig:block_approach2}
     \end{subfigure}
     \hfill
     \begin{subfigure}[b]{0.45\textwidth}
         \centering
         \includegraphics[width=\textwidth]{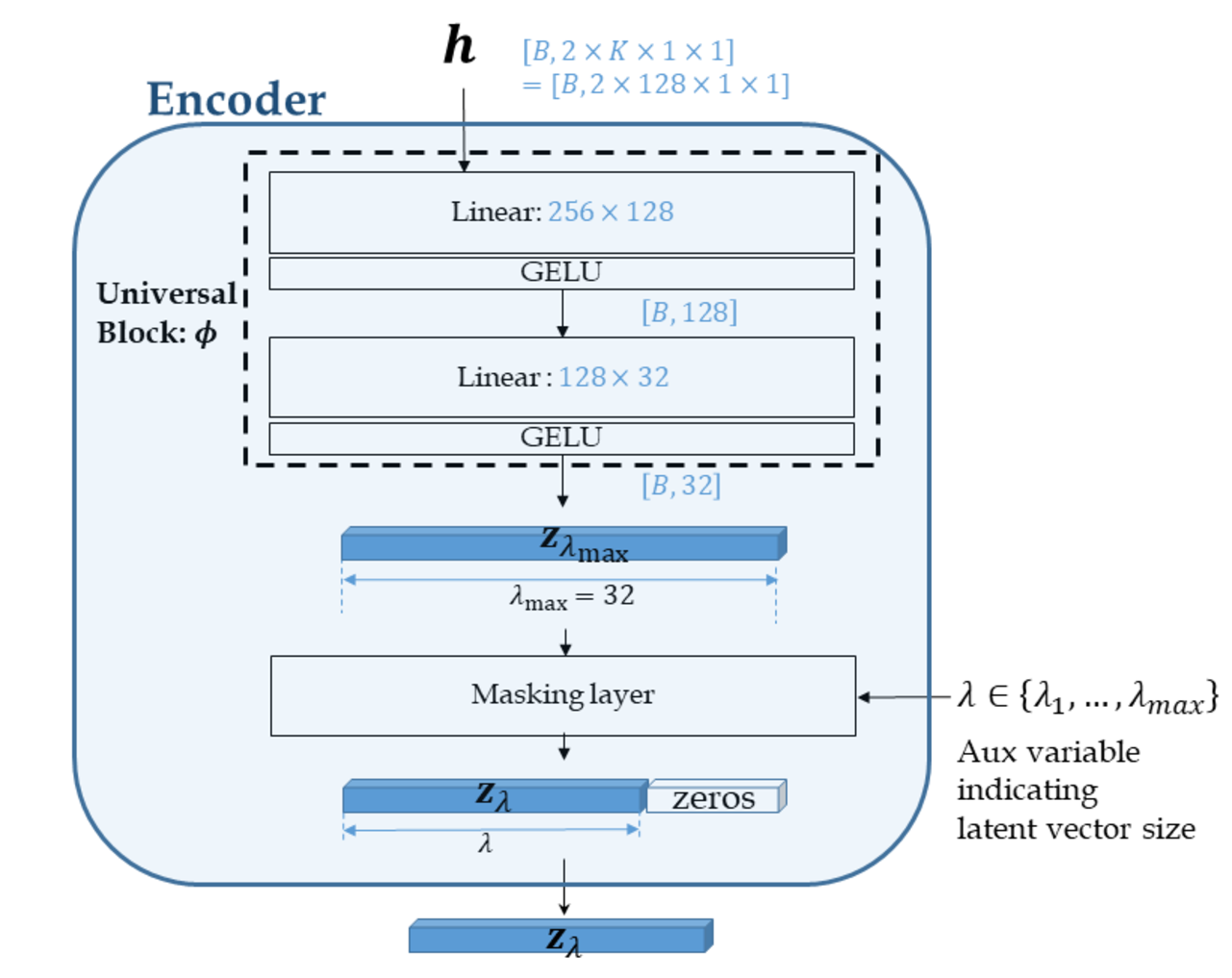}
         \caption{Our proposed approach}
         \label{fig:block_approach3}
     \end{subfigure}
        \caption{Example of block diagram of three approaches to support multiple compression ratios with simple MLP architecture. Universal block in (b) and (c) can be any ML architecture such as CNN or transformer.}
        \label{fig:block_diagram}
\end{figure}

% Before introducing our proposed scheme, we introduce two baseline approaches. 
The first baseline is a naive approach which trains the pair of encoder and decoder for each latent vector size $\lambda_i\in\Lambda$ by minimizing the loss function in \eqref{eq:conv_obj}. Fig.~\ref{fig:block_approach1} shows an example of the MLP architecture using the naive approach to support $\Lambda=\{\lambda_1,\lambda_2,\lambda_3,\lambda_4\}=\{4,8,16,32\}$. This is the worst scheme in terms of the number of parameters. 

The second baseline is introduced in \cite{song2021saldr} which proposes a multiple compression ratio network, named SALDR. It consists of a universal block and serial compression blocks, named fully connected block (FCB), as shown in Fig.~\ref{fig:block_approach2}. 
The universal block first extracts CSI features of size $\lambda_{max}=\lambda_4$. 
Then, FCBs sequentially compress the latent vectors with smaller compression ratios. Employing the universal block can significantly save the HW complexity compared to the naive approach, but it has two limitations. 
First, as the fully connected layer is sequentially applied, latency to generate the latent vector of the smallest size can be very large during the inference phase. 
Second, this architecture is difficult to be extended to the case that the cardinality of $\Lambda$ is large, i.e., the encoder is required to support many cases of compression ratios. 
This is because one additional element in $\Lambda$ requires an additional FCB.

To address these challenges, we propose a new architecture as shown in Fig.~\ref{fig:block_approach3}. It consists of two parts, universal encoding block and a masking layer. The universal encoding block first extracts CSI features of size $\lambda_{max}$ which denotes the largest latent vector size that the AE should support. Then, the masking layer bypasses the first $\lambda$ elements of the output of the universal block and makes the other elements zeros, i.e., auxiliary integer variable $\lambda$ determines the size of the encoder output. 
The key idea is that parameters of the universal block are trained such that the element in the earlier position of the latent vector $\mathbf{z}_{\lambda_{max}}$ contains more important information than elements in the later position. 
This enables our architecture naturally select the latent elements from the earlier positions according to the required latent size $\lambda$.
Our architecture has smaller HW complexity and shorter inference latency as we do not require additional layers for each $\lambda\in\Lambda$. For doing so, we first define the loss function as 
\begin{equation}
    D_{\phi, \theta_\lambda} (\lambda) = \mathbb{E}_{\mathbf{h}} \left[ \left\| \mathbf{h} - g_{\theta_{\lambda}}(f_\phi(\mathbf{h}) \odot \mathbf{e}_\lambda)\right\|^2_2 \right],
\end{equation}
where $\mathbf{e}_\lambda \in \{0,1\}^{\lambda_{max}}$ is a binary vector whose first $\lambda$ elements are $1$ while the others are $0$, and $\odot$ denotes element-wise product. Then the total loss function is defined as
\begin{equation} \label{eq:dedicated_loss}
    D_{\phi, \{\theta_{\lambda}\}_{\lambda\in\Lambda}} (\Lambda) = \sum_{\lambda\in\Lambda} w_\lambda D_{\phi, \theta_{\lambda}} (\lambda),
\end{equation}
where $\Lambda$ is a set of latent vector size that we need to support. $\{w_\lambda\}_{\lambda\in\Lambda}$ are weight coefficients to present the importance of loss function associated with $\lambda$ and $\sum_{\lambda\in\Lambda} w_\lambda = 1$. 
The parameters of encoder and decoder are trained to minimize the total loss function $D_{\phi, \theta} (\Lambda)$, which can be expressed by
\begin{equation} \label{eq:final_obj_function}
    \phi^{\mathtt{opt}}_{\mathtt{univ}}, \{ \theta_\lambda^{\mathtt{opt}} \}_{\lambda\in\Lambda} 
    = \underset{\phi, \{ \theta_\lambda \}_{\lambda\in\Lambda} }{\arg\max} \sum_{\lambda\in\Lambda} w_\lambda D_{\phi, \theta_{\lambda}} (\lambda).
\end{equation}
Note that $\{w_\lambda\}_{\lambda\in\Lambda}$ plays an important role for the distribution of reconstruction loss over $\lambda\in\Lambda$, and we find the values via hyper-parameter tuning. Reinforcement learning can be utilized to optimize it, which would be one of the interesting future directions.

% \begin{figure}[t]
% \centering
% \includegraphics[width=0.9\linewidth]{figures/block_approach1.eps}
% \caption{Objectives}
% \label{fig:block_approach1}
% \vspace{-0.3cm}
% \end{figure}

% \begin{figure}[t]
% \centering
% \includegraphics[width=0.7\linewidth]{figures/block_approach2.eps}
% \caption{Objectives}
% \label{fig:block_approach2}
% \vspace{-0.3cm}
% \end{figure}

\begin{figure}[t]
\centering
\includegraphics[width=\linewidth]{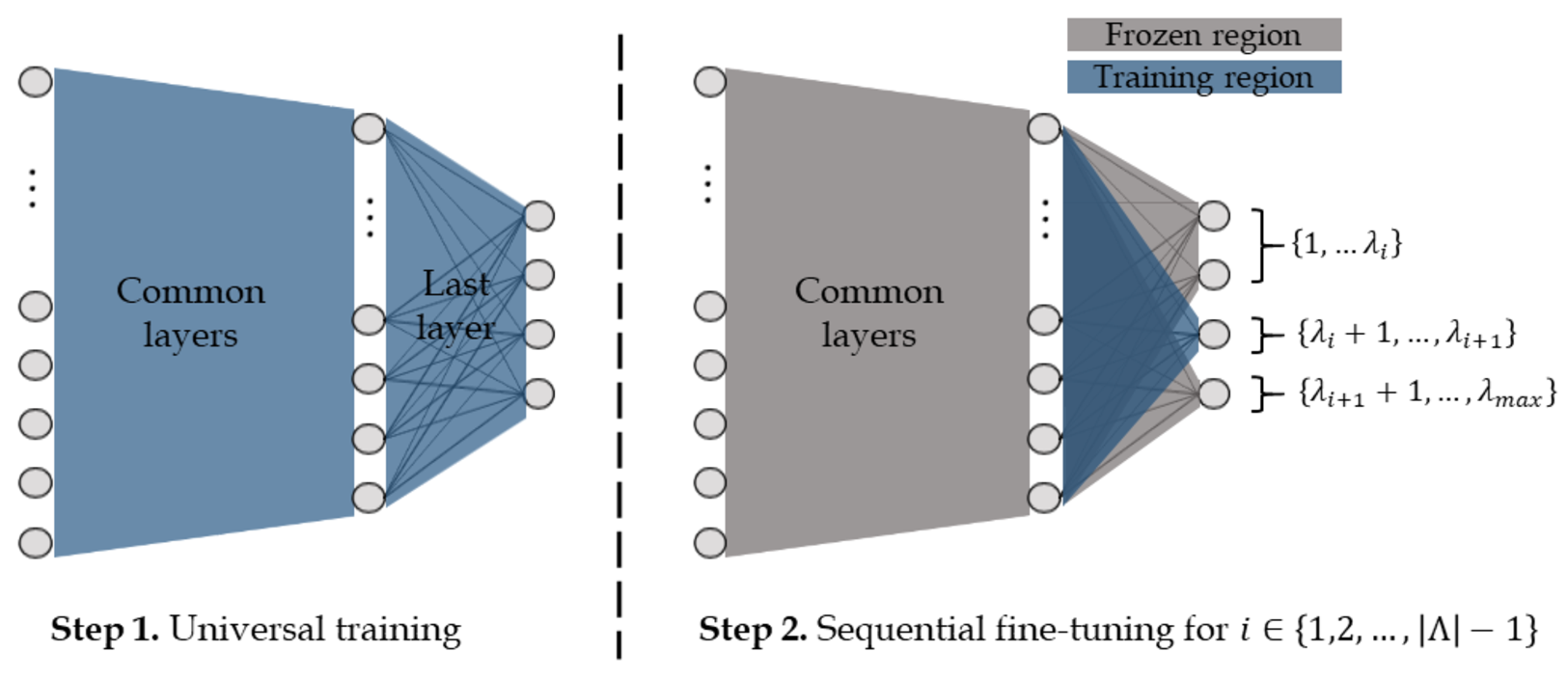}
\caption{Two-step training strategy for fine-tuning.}
\label{fig:fine_tuning}
\vspace{-0.1cm}
\end{figure}

We further propose a two-step training strategy to improve the reconstruction performance of each compression ratio by leveraging the idea of freezing and fine-tuning in transfer learning~\cite{weiss2016survey, guo2019spottune}. The universal block can be divided into two parts, common layers and the last layer, as depicted in Fig.~\ref{fig:fine_tuning}. Key intuition behind the two-step training strategy is that parameters of the common layers contribute to the all compression ratios while parameters of the last layer (commonly fully-connected layer) contribute to a certain compression ratio. In the first step, entire parameters in the common and last layers are trained with the objective function of \eqref{eq:final_obj_function}. In the second step, the parameters of the common layers are frozen and partial parameters of the last layer are sequentially trained with $|\Lambda|$ sub-steps. 
In $i$-th sub-step ($i\in\{1,\ldots,|\Lambda|\}$), the parameters of the last layer linked to elements of the latent vector with element indexes in $\{\lambda_{i-1}+1,\ldots,\lambda_{i}\}$ are trained to minimize the reconstruction loss of $D_{\phi, \theta_{\lambda_i}}(\lambda_i)$ in \eqref{eq:dedicated_loss} while the other parameters are frozen. 
We observe that this two-step training strategy is especially effective to complicated ML architectures such as transformer-based network \cite{mourya2022spatially}, which will be detailed in Section \ref{sec:exp}.
% Due to the limited space, experimental results of fine-tuning the transformer-based network will be included in the final version of the paper.
% , which will be presented in Section \ref{sec:exp}.

\section{Simulation Results}\label{sec:exp}
In this section, we empirically demonstrate that our proposed AE framework has comparable performance in terms of distortion-compression ratio trade-off while it significantly reduces the HW and time complexity in the UE. 

\begin{table}[t!]
\small
\vspace{0.3cm}
\caption{Summary of settings to generate dataset.}
\vspace{-0.2cm}
\label{tbl:dataset}
\begin{center}
% \begin{small}
% \begin{sc}
\begin{tabular}{ccc}
\toprule
  & Settings & \# settings.    \\
\midrule
 channel profile & EPA, EVA, TDL~\cite{3gpp38901tr} & $8$ \\
 SNR (dB) & $11,12,\ldots,40$    & $40$\\
 $K$      & $68,72,\ldots,128$ & $16$ \\
 BS antenna index &  $1,2,\ldots,32$ & $32$\\
 UE antenna index &  $1,2,3,4$ & $4$\\
 \bottomrule
\end{tabular}
% \end{sc}
% \end{small}
\end{center}
\vspace{-0.3cm}
% \vskip -0.1in
\end{table}

\noindent \textbf{Dataset and pre-processing.} 
For a training and test dataset, we generate channel tensors according to the 3rd Generation Partnership Project (3GPP) standard \cite{3gpp38901tr} with various resource block size (up to $K=128$), the number of BS and UE antennas (up to $N_{BS}=32$, $N_{UE}=4$), signal to noise ratio (SNR), and channel profile. 
% We consider the number of subcarrier $K$ in the 4-th category defined in Table~\ref{tbl:input_category}.
As described in Section \ref{subsec:input_generalization}, the generated CSI tensor is partitioned such that each part has a single element in the antenna dimensions, and then each part is zero-padded and transformed to delay domain.
% then scaled such that minimum and maximum value over the entire element in the training dataset becomes zero and one, respectively.  
The detailed settings are summarized in Table 2, and we randomly generated $12$ samples per setting, which results in $5,\!898,\!240$ samples in total.

\noindent \textbf{Implementations.}
For the performance evaluation, we implement three approaches that support four latent sizes $\lambda_1=4$, $\lambda_2=8$, $\lambda_3=16$, and $\lambda_4=\lambda_{\mathtt{max}}=32$, which correspond to the compression ratio of $1/64, 1/32, 1/16, 1/8$ with $K=128$, respectively. 
% For the fair comparison, input space generalization described in Section \ref{subsec:input_generalization} is applied to all three approaches.
Three approaches are implemented with MLP and the detailed description is included in Section \ref{subsec:latent_generalization}. 
\begin{itemize}[leftmargin=*, itemsep=1pt]
    \item \textbf{Approach 1 (Naive approach).} For each $\lambda_i$, UE employ and train the encoder ML model $\phi_i$ separately. This approach is the worst in terms of HW complexity. 

    \item \textbf{Approach 2 (SALDR)}~\cite{song2021saldr}. This is the state-of-the-art scheme to support multiple compression ratios, which requires additional fully connected layers to reduce the size of latent vector.

    \item \textbf{Approach 3 (Proposed)}. As the masking layer has no trainable parameter, our proposed approach is the best in terms of the number of parameters (HW complexity) as well as latency (inference time). 
\end{itemize}

\noindent \textbf{Performance comparison.}
Table \ref{tbl:hw_comparison} compares the number of parameters and latency of the three approaches to implement MLP encoder(s) to support multiple latent vector sizes $\Lambda=\{4,8,16,32\}$ and $\Lambda=\{1,2,3,\ldots,32\}$. 
For the latency, we measure the inference time to compress a single CSI tensor $\mathbf{H}\in \mathbb{R}^{2\times K(=128) \times N_{BS} (=32) \times N_{UE} (=4)}$ in the worst scenario. For each compression ratio, we measure time to carry out forward propagation of the encoder $128$ times as $\mathbf{H}$ is partitioned into $128$ parts $\mathbf{h}^\mathtt{part}\in\mathbb{R}^{2\times K \times 1 \times 1}$, and then report the longest inference time of the CR among $\Lambda$. We run $10,000$ times and report the average of running time.

\begin{figure}[t]
\centering
\vspace{-0.1cm}
\includegraphics[width=0.9\linewidth]{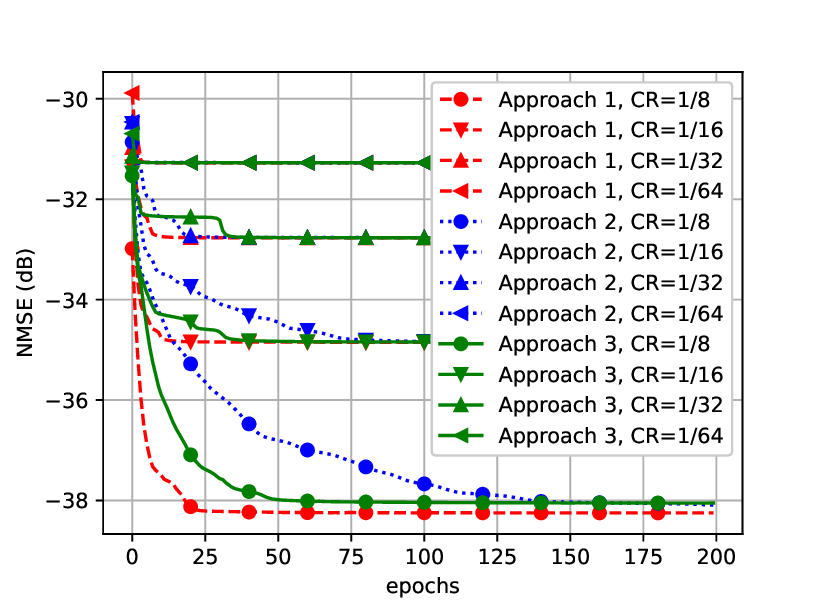}
\caption{NMSE performance of three approaches with MLP.}
\label{fig:nmse}
\vspace{-0.1cm}
\end{figure}

\begin{table}[t!]
\small
\caption{Comparison of HW complexity and latency of three approaches implemented with MLP in two cases with different cardinality of the set of compression ratios, $|\Lambda|$.}
\vspace{-0.2cm}
\label{tbl:hw_comparison}
\begin{center}
% \begin{tabular}{lccc}
% \toprule
%   & \# params. & FLOPs & Latency (ms)   \\
% \midrule
%  Approach 1 (Naive) & $79.84$k & $71.04$k & $0.1470$ \\
%  Approach 2 (SALDR\cite{song2021saldr}) & $37.72$k & $37.53$k & $0.3963$ \\
%  Approach 3 (Ours) & $37.02$k & $36.86$k & $0.1455$ \\

\begin{subtable}[h]{0.45\textwidth}
\begin{tabular}{lccc}
    \toprule
      & \# Params. & Latency (ms)   \\
    \midrule
     Approach 1 (Naive) & $79.84$k & $0.1470$ \\
     Approach 2 (SALDR\cite{song2021saldr}) & $37.72$k  & $0.3963$ \\
     Approach 3 (Ours) & $37.02$k  & $0.1455$ \\
    \bottomrule
    \end{tabular}
    \vspace{0.1cm}
    \caption{Case 1: $|\Lambda|=|\{4,8,16,32\}|=4$,}
   \label{tab:week1}
    \end{subtable}
\hfill
\begin{subtable}[h]{0.45\textwidth}
\begin{tabular}{lccc}
    \toprule
    & \# Params. & Latency (ms)   \\
    \midrule
    Approach 1 (Naive) & $589.1$k & $0.1446$ \\
    Approach 2 (SALDR\cite{song2021saldr}) & $48.43$k  & $2.708$ \\
    Approach 3 (Ours) & $37.02$k  & $0.1467$ \\
    \bottomrule
    \end{tabular}
    \vspace{0.1cm}
    \caption{Case 2: $|\Lambda|=|\{1,2,3,\ldots,32\}|=32$.}
    \label{tab:week2}
    \end{subtable}
\end{center}
\vspace{-0.5cm}
\end{table}

\begin{figure}[!t]
\centering
\vspace{-0.1cm}
\includegraphics[width=0.8\linewidth]{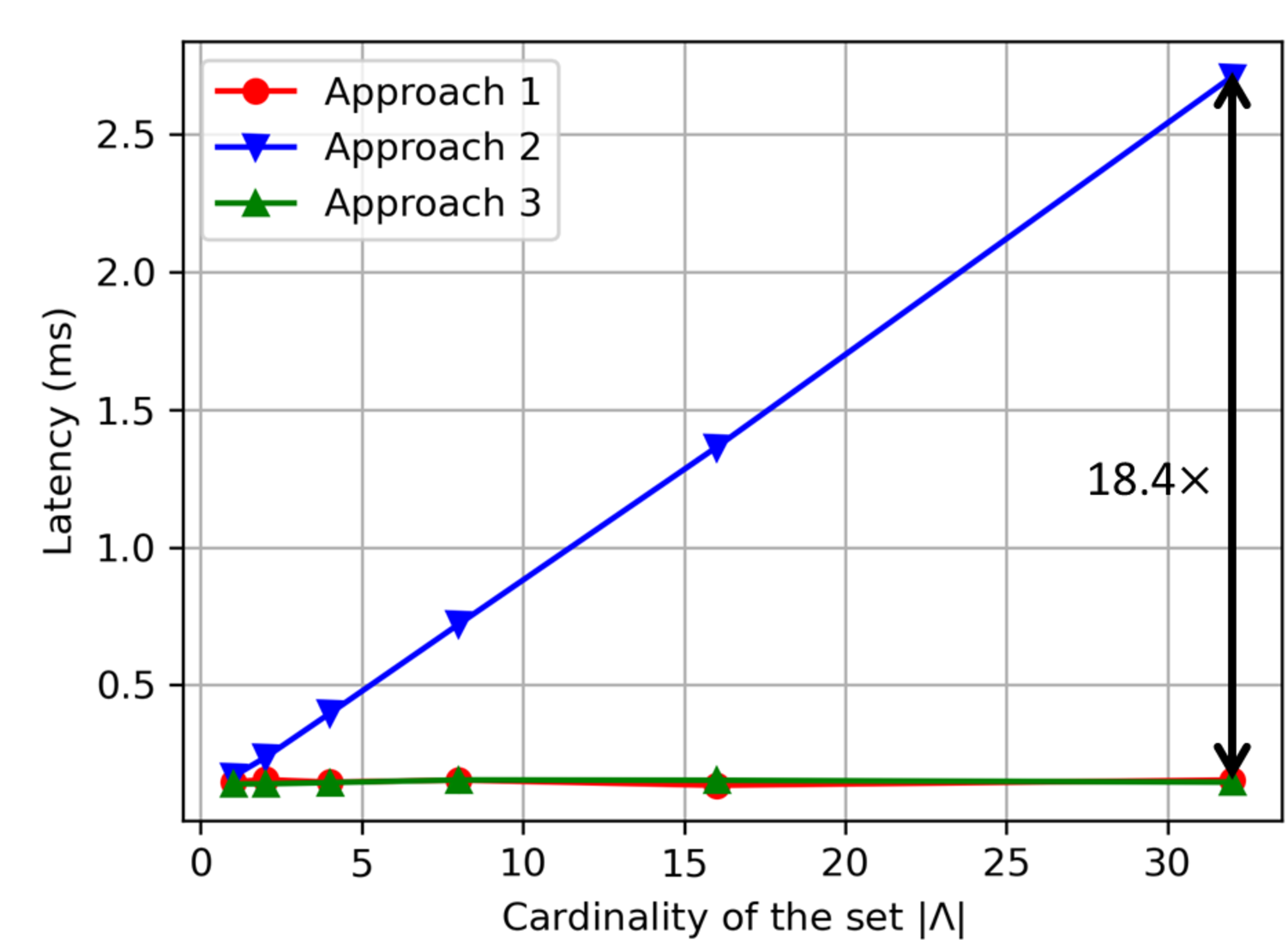}
\caption{ Inference time (latency) of three approaches with the increasing of the cardinality of the set $| \Lambda |$.}
\label{fig:latency}
\vspace{-0.1cm}
\end{figure}

\begin{figure}[!t]
\centering
\vspace{-0.1cm}
\includegraphics[width=0.8\linewidth]{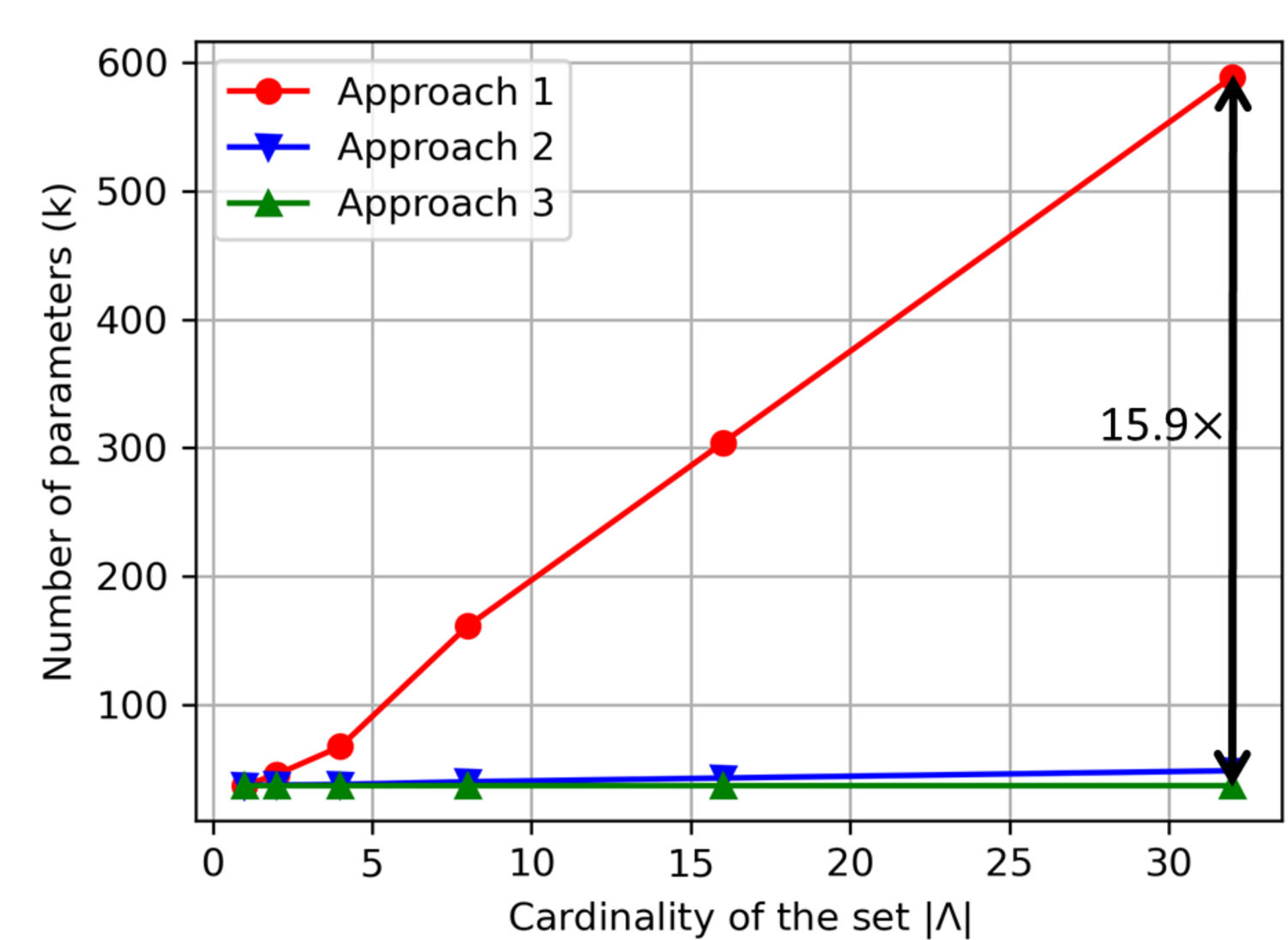}
\caption{ The number of parameters of three approaches with the increasing of the cardinality of the set $| \Lambda |$.}
\label{fig:params}
\vspace{-0.1cm}
\end{figure}

% Figure 10 and 11 show the NMSE ($\| \mathbf{h} - \hat{\mathbf{h}}\|^2_2 / \| \mathbf{h}\|^2_2$) and BLER performance of the three approaches, respectively. 
Fig.~\ref{fig:nmse} shows the NMSE ($\| \mathbf{h} - \hat{\mathbf{h}}\|^2_2 / \| \mathbf{h}\|^2_2$) performance of the three approaches with four different compression ratios (CRs).
Fig.~\ref{fig:latency} and Fig.~\ref{fig:params} shows the latency and the number of parameters of the three approaches with increasing the cardinality of the set $\Lambda$, respectively.
We have the following observations:
\begin{itemize}[leftmargin=*, itemsep=1pt]
    \item Three approaches have the almost same NMSE performance while the proposed approach 3 has much smaller storage and computational complexity than the Approach 1 and 2.

    \item Our proposed approach reduces the number parameters comparing to Approach 1 (up to $15.9\times$) and Approach 2 (up to $1.3\times$). This gain becomes larger when the cardinality of the set $|\Lambda|$ becomes larger as shown in Fig.~\ref{fig:params}.

    \item Our proposed approach significantly reduces the inference time (latency) comparing to Approach 2 because Approach 2 sequentially applies the fully connected layers to reduce the latent vector size. This latency gap becomes larger when the cardinality of the set $|\Lambda|$ becomes larger as shown in Fig.~\ref{fig:latency}.

    \item One of key contributions of the proposed approach is that HW storage complexity and latency is independent on the cardinality of the set $|\Lambda|$.
    % while those of approach 1 and 2 linearly increases as the cardinality of the set $|\Lambda|$ increases.     
\end{itemize}

\begin{figure}[!ht]
\centering
\vspace{-0.1cm}
\includegraphics[width=0.95\linewidth]{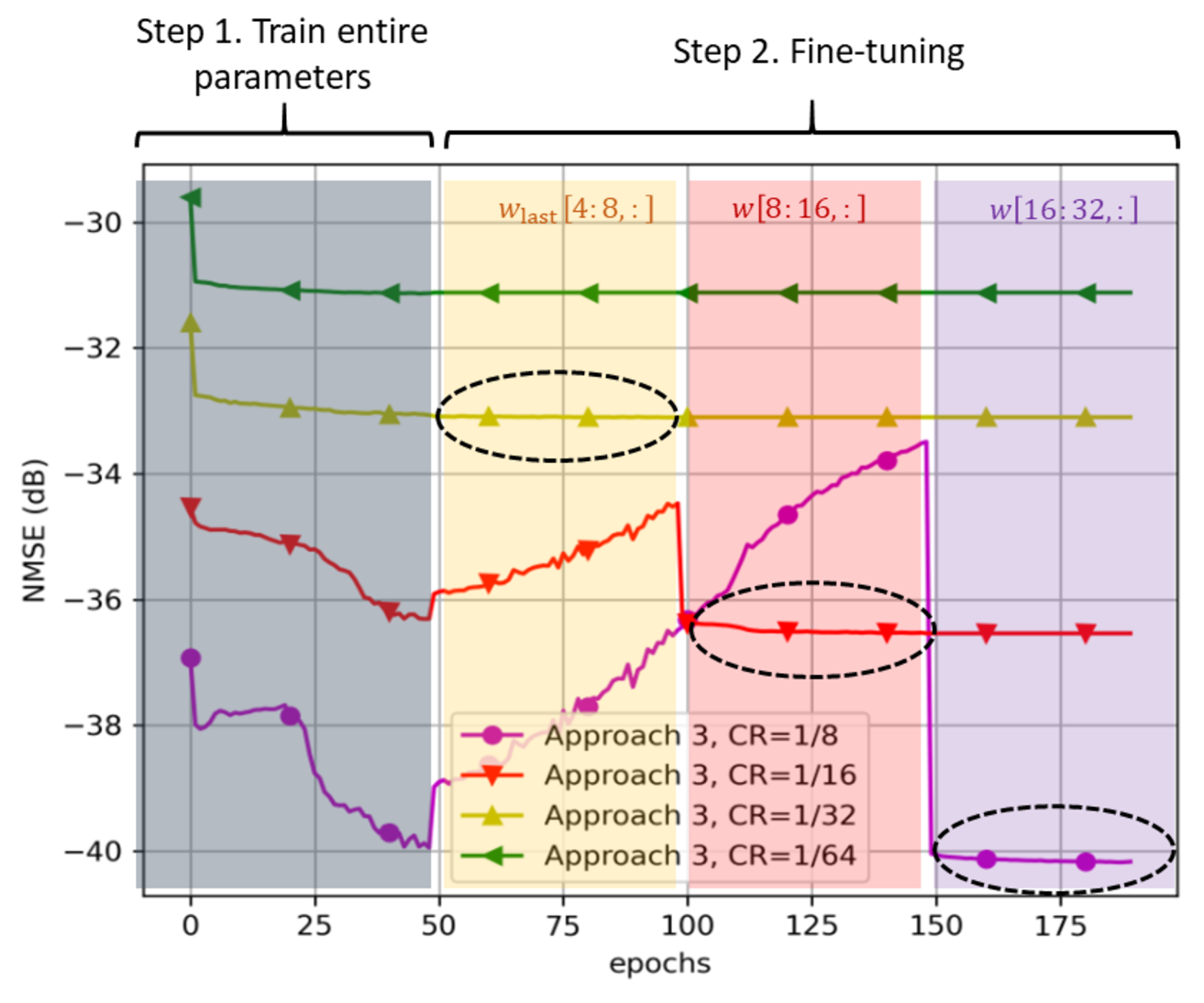}
\caption{ Learning curve of our proposed approach with two-step training strategy. Transformer-based architecture in \cite{mourya2022spatially} is implemented for the universal block depicted in Fig.~\ref{fig:block_approach3}. }
\label{fig:fine-tuning-results}
% \vspace{-0.1cm}
\end{figure}

\noindent\textbf{Two-step training strategy.}
Universal block in the proposed approach can be any ML architecture. When the complicated ML architecture such as a transformer-based network\cite{mourya2022spatially} is implemented for the universal block, we observe that learning curve with lower compression ratio (e.g., CR=$1/64$) fluctuates more than the higher CR (e.g., CR=$1/8$). Two-step training strategy described in Section~\ref{subsec:latent_generalization} can address this problem. Fig.~\ref{fig:fine-tuning-results} shows the learning curve of the proposed approach with the two-step training strategy. We can observe that in sub-step of the fine-tuning step, learning curve of the target CR is properly trained while the learning curve of the higher CR remains constant. For instance, in the second sub-step where the target CR is $1/16$ (epochs $\in [100,150)$), we train the parameters in the last layer of the universal block connected to latent element indexes in $[8,16)$. Parameters connected to the latent element indexes in $[0,8)$ is frozen as they have impact on the performance of lower CRs, $1/32$ and $1/64$. Block dotted circles in Fig.~\ref{fig:fine-tuning-results} indicate the learning curve of the target CR at each sub-step of the fine-tuning step.

\section{Conclusion}
In this paper, we consider AE-based CSI compression framework with continuous representation in the latent vector. In practice, however, the feedback link has the constraint capability of conveying only finite bits. Investigating how to apply quantization techniques such as \cite{ravula2021deep, kong2021knowledge} to our framework would be an interesting future direction. 

% Future directions:
% \begin{itemize}
%     \item \textbf{Quantization over latent space.} In this paper, we consider dimensionality reduction assuming that elements of latent vector have the continuous value. In practice, however, the feedback link has the constraint capability of conveying only finite bits. Investigating how to apply quantization techniques in \cite{ravula2021deep, kong2021knowledge} to our framework would be an interesting future direction. 
    
% \end{itemize}

\bibliographystyle{IEEEtran}
\bibliography{main.bbl}

\end{document}